\documentstyle[11pt,aaspp4,flushrt]{article} 
\begin{document}

\title{The Spectra of Main Sequence Stars in Galactic Globular Clusters
II. CH and CN Bands in M71\altaffilmark{1}}

\author{Judith G. Cohen\altaffilmark{2}}

\altaffiltext{1}{Based on observations obtained at the
W.M. Keck Observatory, which is operated jointly by the California 
Institute of Technology and the University of California}
\altaffiltext{2}{Palomar Observatory, Mail Stop 105-24,
California Institute of Technology}

\begin{abstract}

Spectra with a high signal-to-noise ratio of 79 stars 
which are just below the main sequence
turnoff of M71 are presented.  They yield
indices for the strength of the G band of CH and the ultraviolet CN band 
at 3885 \AA.
These indices are each to first order bimodal and they are anti-correlated.
There are approximately equal numbers of CN weak/CH strong and
CN strong/CH weak main sequence stars in M71.
It is not yet clear whether these star-to-star
variations arise from primordial variations or from
mixing within a fraction of individual stars as they
evolve.

\end{abstract}


\subjectheadings{globular clusters: general --- globular clusters: individual (M71) --- stars: evolution}

\section{INTRODUCTION}

The major issue I intend to explore in this series of papers
is that of star-to-star
variations in abundances within a single globular cluster
at and below the level of the main sequence turnoff.
In the first paper of this series (Cohen 1999), I presented
an overview of this subject and the historical background
for the present work which began with the study of the giants in
M13 by Suntzeff (1981).  To summarize very briefly, star-to-star
variations in several light elements, particularly C and N,
are seen on the giant and subgiant branches of many globular
clusters.  The behavior of these variations within those
globular clusters studied in most detail are in many cases consistent
with what is expected for mixing as the explanation for most of the observed
variations.  47 Tuc was the only galactic globular cluster
studied at the level of the main sequence, where mixing should
not yet have occurred, yet Canon {\it{et al.}} (1998) and references therein
find detectable variations in the CH and CN bands for main
sequence stars in 47 Tuc, and these variations are anti-correlated.

Globular cluster main sequence stars should
not yet have synthesized through internal nuclear burning any elements
heavier than He (and Li and Be) and hence will
be essentially unpolluted by the internal nuclear burning 
and production of various heavy elements
that occur in later
stages of stellar evolution.  Theory predicts that these stars are
unaffected by gravitational settling and that their surfaces should
be a fair representation of the gas from which the globular cluster formed.
Thus the persistence of variations in C and N 
to such low luminosities in 47 Tuc (Cannon {\it{et al.}} 1998) and
references therein) is surprising.  The major recent reviews in this area
are those of Kraft (1994) and Briley, Hesser \& Smith (1994), while the
more general reviews of  
McWilliam (1997) and of Pinnsoneault (1997) are also relevant.

In paper 1 I analyzed spectra of 50 stars at or below the main sequence 
turnoff in M13.  I did not find any detectable variation
in the strength of the CN or CH features in this sample.
However, M13 is a quite metal poor globular cluster, and hence here
I present an analysis of the spectra of 79 stars on the main sequence
of M71, a 
globular cluster with metallicity comparable to that of 47 Tuc.

\section{THE SAMPLE OF STARS}

M71 was chosen as the second cluster in this study because of its
metallicity (Cohen 1983) and because
it is nearby, hence the
turnoff stars will be relatively bright.
This globular cluster is located at galactic latitude bII = $-4.6^{\circ}$
and has a reddening $E(B-V) = 0.27$ mag (Reed, Hesser \& Shawl 1988).

Short exposure images in $B$ and $R$ were taken with Low Resolution
Imaging Spectrograph (Oke {\it{et al.}} 1995) at the Keck
Observatory centered on the cluster. 
Photometry was obtained 
with DAOPHOT (Stetson 1987) using these short exposures
calibrated on the system of Landolt (1992).
The zero point for each color is uncertain by $\pm0.05$ mag.
A sample of main sequence stars was chosen based on their position
on the locus of the main sequence as defined by this photometry.
Each candidate was inspected for crowding and stars were chosen
for the spectroscopic sample on the
basis of minimum crowding. Table 1 gives the object's 
coordinates (B1950), $R$ mag,
$B-R$ color, and indices (together with their errors) for two molecular bands,
the G band of CH at 4305 \AA\ and the CN band near 3880 \AA,
for the M71 main sequence stars in the spectroscopic sample.
The magnitudes and colors given in Table 1
have not been corrected for extinction.

Since the fields are very crowded, in addition to providing
the star coordinates, we provide an identification chart (Figure 1) for
a few stars in the M71 sample, from which, given the
accurate relative coordinates, the rest of the stars can be located.
Relative stellar coordinates are defined from the LRIS images themselves
assuming the telescope pointing recorded in the image header is correct
and taking into account the optical distortions in the images.  The
astrometry of Cudworth (1985) is used to fix the
absolute coordinates.

%
%
%

Figure 2 presents a
color-magnitude diagram for the main sequence stars in the M71 sample.  
The stars that have been observed spectroscopically are displaced
by $-0.6$ mag in $B-R$ color and are shown
as filled circles.  

%

The color-magnitude diagram of the field of M71 shown
in Figure 2 reflects the low galactic latitude this cluster.
There are many field stars, most redder and presumably more distant
than the globular cluster itself.
Because the radial velocity of M71 is $-27$ km s$^{-1}$ (Cohen 1980),
it is not possible to isolate a sample of cluster members using radial velocity
measurements from low dispersion spectra.  My sample will therefore
have some field star contamination.  To estimate the fraction of
field stars in such a sample, Figure 3 shows a histogram in color
of all stars with $17.3 < R < 17.6$ mag.  The sharp rise at the
blue end of the distribution arises from the cluster, while the
extended tail to the red is from field stars.  

The M13 main sequence shown in Paper 1
is very narrow in color.  Some of the
apparent spread in color seen in the case of M71 may be due to
variations in reddening.  These are easily detected at the level of
$\sim25\%$ of E$(B-V)$  (and corresponding amounts in other colors)
from multicolor photometry of stars on the red giant branch
in more heavily reddened galactic globular clusters
(Cohen \& Sleeper 1995).  In addition, the field of M71 is very crowded, even
more so than either of the two fields in M13 studied in Paper 1.
This may lead to photometric errors which might produce 
an apparent spread in color of the M71 main sequence.

Based on Figure 3
one might estimate a field star contamination which at worst does not exceed
25\% of the sample.

%

\section{SPECTROSCOPIC OBSERVATIONS AND MEASUREMENT OF BAND INDICES}

Three slitmasks were 
designed containing 79 stars from the M71 main sequence
star sample.  These were used at relatively low dispersion
with the LRIS (300 g/mm grating, 2.46\AA/pixel, 0.7 arcsec slit width) for
a spectral resolution of 8\AA.  The CCD detector is digitized
at 2 electrons/DN with a readout noise of 8 electrons.
Two 800 sec exposures were
taken with each slitmask under conditions of good seeing and
dark sky in the summer of 1998.  The data were reduced in
a straightforward manner as described in Cohen {\it{et al.}} (1999)
using Figaro (Shortridge 1988) except that the wavelength calibration came
from arc lamp exposures, rather than from night sky lines on the
spectra themselves.
The spectra are not fluxed.

The definition of the CH and uvCN indices follows that of Paper 1,
except that the wavelengths are shifted to take into account
the difference in mean radial velocity between M13 and M71.
The CH index again uses
continuum bandpasses on both sides of the G band at 4305\AA, with
a feature bandpass chosen to avoid H$\gamma$.  
The CH and uvCN indices thus measured, together with their $1\sigma$ 
errors calculated assuming Poisson statistics are given in Table 1.  
The values are the fraction of absorption from the continuum,
and are not in magnitudes.  Recall that the errors given in Table 1 do
not include the effect of cosmic rays nor the effect of the 
background signal from the night sky, both of which are small.

The M71 main sequence stars in my sample are somewhat brighter than those
of M13, and the spectra are thus of even higher signal-to-noise than those
of my main sequence sample in M13.  Illustrative examples of the latter
are shown in Figure 3 of Paper 1.

The continuum level in one star of the 79 in the M71 
main sequence sample fell slightly below the
minimum value (700 DN/pixel) set in Paper 1 for accepting the
uvCN index measurements.  However the features are so much
stronger in this globular cluster that even the value for this object was accepted.

\section{ANALYSIS}

Figure 4 shows the CH index plotted as a function of $R$ mag for the
79 stars in my M71 main sequence sample.  The 1$\sigma$ error
bars are shown for each star.  It is immediately clear that there
is a large range in CH strength at all luminosities, which range is
many times the measurement uncertainties.

Four of the stars in my M71 sample have very
strong CH for their magnitude and are believed to be field stars.
They are indicated by ``x'' symbols and are the only objects
within the rectangle at the upper right of the figure.
Ignoring these four stars, a second order least squares fit was
carried out of CH index as a function of $R$ mag.  Objects that lie above
the mean fit are shown in Figure 4 as open circles, while objects
that lie below this curve are shown as filled circles.

Ignoring the four probable field stars, the 
distribution of CH indices now appears to be approximately bimodal for
the M71 sample.

Figure 5 show the results for the uvCN indices in the 79 M71 main sequence
stars, again plotted as a function of $R$ mag.  As was the case for
the CH band indices, a large range
in the strength of the uvCN band at a fixed luminosity is seen for 
the M71 main sequence. The overall appearance of the distribution
is that it is bimodal.  The same symbols
are used in Figure 5
as in Figure 4.  Comparing the two figures, it is immediately
apparent that the CH and uvCN indices are anti-correlated.
There are approximately equal numbers of CH strong/CN weak and CH weak/CN strong
stars.  Langer, Suntzeff \& Kraft (1992) find this fraction determined
from spectra of the red giants to vary
from cluster to cluster among a set of three galactic globular
clusters (M3, M13 and M79) of intermediate metallicity.

The good correlations seen in Figures 4 and 5 provide evidence
that my M71 main sequence sample is not seriously contaminated by field stars.

%
%

\section{DISCUSSION}

In Paper 1 I showed that my 
data provide no evidence for variations of CH or CN band strengths
among the 50 main sequence stars in our sample in M13.  Here in the
case of M71 I have found a range in CH and uvCN indices at a given
luminosity along the M71 main sequence which is much larger than the
measurement errors.  Furthermore, the CH and uvCN indices are
anti-correlated, and both appear to be bimodal.

It may well be that even though the measurement errors in M13 are small,
the low metallicity of the cluster and the definition of
such relatively crude molecular band indices conspire to hide any
variation that may actually be present.  The next paper in this
series (Briley \& Cohen 1999) will explore this possibility and
will attempt to provide a guess as to the range of C and N variations
that may be present in each of the two globular clusters, M71 and M13.

The next step will be to analyze other light elements in M71 to
see if variations along the main sequence can be detected in Na, Ca or 
Mg, for example.  Na and to a lesser extent Mg are known to vary 
among the giants and subgiants in several globular clusters.  
Suitable data from the
LRIS at the Keck Observatory consisting of spectra of
significantly higher dispersion than those analyzed here or in Paper 1
are already in hand for both the M71
and the M13 sample.  These spectra
can also be used to eliminate at least some of the
field stars in the M71 sample through
radial velocities.

Now that it is clear that variations of C and N are strong in
at least two metal rich galactic globular clusters at the level of the
main sequence, a major effort needs to be mounted to differentiate
between mixing and primordial variations.  Understanding the
origin of these star-to-star variations at the level of the main sequence
is an issue of importance
not only to the field of globular cluster studies, but also has
ramifications throughout many areas of stellar evolution and
galaxy halo ages. I will return to this
issue in future papers in this series.

\section{SUMMARY}

I have determined the strength of the CH and CN bands 
from spectra of 79 main sequence stars in M71. 
Significant variations in the strength of the G band of CH 
at 4305 \AA\ and of
the ultraviolet CN band at 3885 \AA\
are seen from star to star at a fixed luminosity on the main
sequence of M71.  Both the CH indices and the uvCN indices appear
to be bimodal and they are anti-correlated.
There are approximately equal numbers of CN weak/CH strong and
CN strong/CH weak main sequence stars in M71.

This is in contrast to the case of M13 discussed in Paper 1, where no
variations are 
seen for C and N in M13 at the level of the main sequence turnoff
and below it.  I suggest that the variations may actually be present
in M13 but cannot be detected with the molecular band indices
I am using due to  the low metallicity of M13 and to the lack of 
sensitivity of the molecular band indices themselves.
The origin of this behavior, whether it
is due mixing or to primordial variations or to some combination
of these two factors, is not yet clear.

\acknowledgements The entire Keck/LRIS user community owes a huge debt
to Jerry Nelson, Gerry Smith, Bev Oke, and many other people who have
worked to make the Keck Telescope and LRIS a reality and to
operate and maintain the Keck Observatory.  We are grateful
to the W. M. Keck Foundation, and particularly its late president,
Howard Keck, for the vision to fund the construction of the W. M. Keck
Observatory.   I also thank  Kevin Richberg for help with the data
reduction.

\newpage

\begin{deluxetable}{lrrrrrr}
\tablewidth{0pt}
\scriptsize
\tablecaption{Properties of the Sample of Main Sequence Stars in M71}
\tablehead{
\colhead{ID}  & \colhead {$R$} & 
\colhead{$(B-R)$} & \colhead{$I(CH)$} & \colhead{$\sigma$(CH)}
& \colhead{$I(uvCN)$} & \colhead{$\sigma$(uvCN)} \nl
 &  \colhead{(mag)} &
\colhead{(mag)} & \colhead{(\%)} & \colhead{(\%)} & 
\colhead{(\%)} & \colhead{(\%)} \nl
}
\startdata
%
%
 M71ms 1951272+183732 &   17.12 &  1.29 &   0.158  &    0.004    &   0.165  &    0.010  \nl
 M71ms 1951215+183720 &   17.00 &  1.38 &   0.167  &    0.005    &   0.165  &    0.009  \nl
 M71ms 1951216+183611 &   17.52 &  1.28 &   0.078  &    0.003    &   0.205  &    0.008  \nl
 M71ms 1951222+183914 &   17.47 &  1.34 &   0.109  &    0.003    &   0.252  &    0.009  \nl
 M71ms 1951224+183638 &   17.60 &  1.39 &   0.086  &    0.004    &   0.225  &    0.010  \nl
 M71ms 1951237+183746 &   17.17 &  1.39 &   0.174  &    0.005    &   0.136  &    0.008  \nl
 M71ms 1951252+183722 &   17.01 &  1.31 &   0.152  &    0.005    &   0.160  &    0.009  \nl
 M71ms 1951257+183914 &   17.21 &  1.40 &   0.114  &    0.004    &   0.260  &    0.008  \nl
 M71ms 1951262+183856 &   17.57 &  1.32 &   0.152  &    0.005    &   0.149  &    0.012  \nl
 M71ms 1951266+183825 &   17.39 &  1.28 &   0.130  &    0.004    &   0.166  &    0.013  \nl
 M71ms 1951266+183940 &   17.35 &  1.36 &   0.135  &    0.004    &   0.178  &    0.010  \nl
 M71ms 1951268+183621 &   17.37 &  1.28 &   0.149  &    0.004    &   0.145  &    0.011  \nl
 M71ms 1951274+183641 &   17.44 &  1.30 &   0.117  &    0.003    &   0.178  &    0.007  \nl
 M71ms 1951274+184033 &   17.35 &  1.38 &   0.161  &    0.005    &   0.176  &    0.011  \nl
 M71ms 1951275+184023 &   17.53 &  1.30 &   0.132  &    0.004    &   0.176  &    0.010  \nl
 M71ms 1951276+183751 &   17.49 &  1.23 &   0.151  &    0.004    &   0.139  &    0.009  \nl
 M71ms 1951279+183947 &   17.54 &  1.42 &   0.169  &    0.006    &   0.155  &    0.013  \nl
 M71ms 1951286+184144 &   17.57 &  1.28 &   0.103  &    0.003    &   0.232  &    0.009  \nl
 M71ms 1951302+184144 &   17.44 &  1.30 &   0.150  &    0.005    &   0.141  &    0.011  \nl
 M71ms 1951303+183943 &   17.26 &  1.29 &   0.086  &    0.003    &   0.253  &    0.008  \nl
 M71ms 1951318+184025 &   17.38 &  1.28 &   0.142  &    0.005    &   0.146  &    0.011  \nl
 M71ms 1951339+184038 &   17.15 &  1.31 &   0.163  &    0.004    &   0.147  &    0.007  \nl
 M71ms 1951365+183958 &   17.25 &  1.34 &   0.151  &    0.004    &   0.145  &    0.006  \nl
 M71ms 1951380+184103 &   17.52 &  1.27 &   0.165  &    0.004    &   0.185  &    0.005  \nl
 M71ms 1951388+184119 &   17.32 &  1.32 &   0.143  &    0.003    &   0.107  &    0.007  \nl
 M71ms 1951399+184126 &   17.43 &  1.27 &   0.097  &    0.002    &   0.182  &    0.007  \nl
 M71ms 1951231+183740 &   17.47 &  1.35 &   0.142  &    0.005    &   0.142  &    0.013  \nl
 M71ms 1951293+183721 &   17.19 &  1.32 &   0.143  &    0.006    &   0.146  &    0.014  \nl
 M71ms 1951294+183804 &   17.59 &  1.28 &   0.142  &    0.005    &   0.136  &    0.012  \nl
 M71ms 1951295+183748 &   17.26 &  1.27 &   0.123  &    0.005    &   0.255  &    0.009  \nl
 M71ms 1951301+183738 &   17.01 &  1.34 &   0.121  &    0.005    &   0.266  &    0.009  \nl
 M71ms 1951321+183715 &   17.40 &  1.29 &   0.133  &    0.005    &   0.173  &    0.010  \nl
 M71ms 1951325+183607 &   17.55 &  1.28 &   0.137  &    0.004    &   0.154  &    0.009  \nl
 M71ms 1951360+183614 &   17.17 &  1.31 &   0.174  &    0.004    &   0.135  &    0.009  \nl
 M71ms 1951211+184010 &   17.35 &  1.31 &   0.145  &    0.005    &   0.144  &    0.012  \nl
 M71ms 1951239+184109 &   17.39 &  1.39 &   0.099  &    0.003    &   0.243  &    0.009  \nl
 M71ms 1951266+184102 &   17.49 &  1.36 &   0.139  &    0.004    &   0.140  &    0.012  \nl
 M71ms 1951230+184200 &   17.43 &  1.35 &   0.095  &    0.003    &   0.220  &    0.008  \nl
 M71ms 1951239+183906 &   17.46 &  1.29 &   0.098  &    0.003    &   0.251  &    0.008  \nl
 M71ms 1951241+183940 &   17.14 &  1.42 &   0.181  &    0.006    &   0.160  &    0.010  \nl
 M71ms 1951247+184144 &   17.13 &  1.40 &   0.130  &    0.004    &   0.281  &    0.007  \nl
 M71ms 1951248+184155 &   17.36 &  1.32 &   0.141  &    0.004    &   0.175  &    0.009  \nl
 M71ms 1951249+183857 &   17.42 &  1.33 &   0.146  &    0.005    &   0.159  &    0.013  \nl
 M71ms 1951253+183831 &   17.13 &  1.36 &   0.124  &    0.005    &   0.253  &    0.008  \nl
 M71ms 1951253+184132 &   17.24 &  1.34 &   0.123  &    0.005    &   0.244  &    0.008  \nl
 M71ms 1951254+184008 &   17.25 &  1.35 &   0.158  &    0.005    &   0.160  &    0.011  \nl
 M71ms 1951264+183933 &   17.20 &  1.34 &   0.117  &    0.004    &   0.243  &    0.009  \nl
 M71ms 1951277+183627 &   17.33 &  1.36 &   0.203  &    0.008    &   0.314  &    0.012  \nl
 M71ms 1951278+183638 &   17.05 &  1.33 &   0.162  &    0.006    &   0.245  &    0.010  \nl
 M71ms 1951283+183952 &   17.48 &  1.25 &   0.131  &    0.005    &   0.173  &    0.012  \nl
 M71ms 1951297+183832 &   17.39 &  1.29 &   0.146  &    0.006    &   0.136  &    0.011  \nl
 M71ms 1951299+183617 &   17.03 &  1.35 &   0.203  &    0.008    &   0.298  &    0.010  \nl
 M71ms 1951311+183525 &   17.45 &  1.38 &   0.187  &    0.007    &   0.324  &    0.010  \nl
 M71ms 1951323+183552 &   17.55 &  1.23 &   0.099  &    0.004    &   0.230  &    0.008  \nl
 M71ms 1951385+183741 &   17.54 &  1.24 &   0.139  &    0.004    &   0.147  &    0.009  \nl
 M71ms 1951387+183512 &   17.16 &  1.31 &   0.157  &    0.004    &   0.126  &    0.007  \nl
 M71ms 1951389+183610 &   17.55 &  1.30 &   0.141  &    0.004    &   0.145  &    0.007  \nl
 M71ms 1951392+183732 &   17.45 &  1.24 &   0.090  &    0.003    &   0.227  &    0.007  \nl
 M71ms 1951403+183817 &   17.22 &  1.31 &   0.152  &    0.004    &   0.158  &    0.007  \nl
 M71ms 1951411+183806 &   17.15 &  1.31 &   0.098  &    0.003    &   0.231  &    0.008  \nl
 M71ms 1951333+184107 &   17.24 &  1.39 &   0.189  &    0.006    &   0.272  &    0.009  \nl
 M71ms 1951364+183720 &   17.26 &  1.25 &   0.128  &    0.004    &   0.149  &    0.009  \nl
 M71ms 1951364+183749 &   17.42 &  1.29 &   0.133  &    0.005    &   0.137  &    0.010  \nl
 M71ms 1951372+183945 &   17.30 &  1.24 &   0.135  &    0.004    &   0.145  &    0.009  \nl
 M71ms 1951372+184149 &   17.01 &  1.40 &   0.126  &    0.004    &   0.296  &    0.007  \nl
 M71ms 1951373+183642 &   17.21 &  1.28 &   0.152  &    0.005    &   0.138  &    0.007  \nl
 M71ms 1951378+183706 &   17.52 &  1.26 &   0.097  &    0.004    &   0.230  &    0.007  \nl
 M71ms 1951383+184003 &   17.11 &  1.33 &   0.170  &    0.006    &   0.267  &    0.007  \nl
 M71ms 1951391+183901 &   17.25 &  1.35 &   0.140  &    0.004    &   0.140  &    0.010  \nl
 M71ms 1951396+184028 &   17.56 &  1.28 &   0.146  &    0.004    &   0.146  &    0.007  \nl
 M71ms 1951400+184016 &   17.14 &  1.27 &   0.122  &    0.003    &   0.214  &    0.006  \nl
 M71ms 1951403+184120 &   17.51 &  1.39 &   0.186  &    0.005    &   0.216  &    0.008  \nl
 M71ms 1951404+183926 &   17.05 &  1.41 &   0.197  &    0.005    &   0.138  &    0.005  \nl
 M71ms 1951405+184141 &   17.03 &  1.36 &   0.125  &    0.004    &   0.268  &    0.007  \nl
 M71ms 1951406+183826 &   17.59 &  1.29 &   0.108  &    0.003    &   0.226  &    0.005  \nl
 M71ms 1951406+183853 &   17.03 &  1.38 &   0.148  &    0.004    &   0.265  &    0.006  \nl
 M71ms 1951417+183631 &   17.29 &  1.34 &   0.103  &    0.003    &   0.218  &    0.005  \nl
 M71ms 1951355+184057 &   17.18 &  1.31 &   0.094  &    0.003    &   0.227  &    0.008  \nl
\enddata
\end{deluxetable}

\newpage

\begin{figure}
\epsscale{0.7}
\plotone{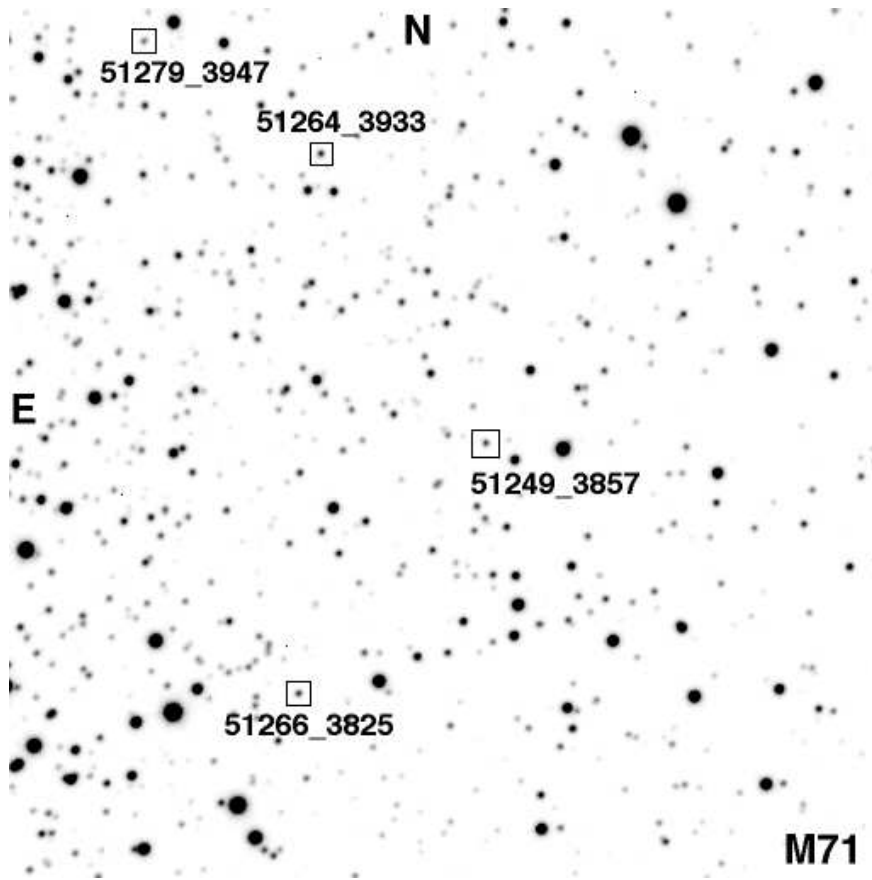}
\caption[jgcohen2.fig1.ps]{A square section 110 arcsec on a side located
slightly West of the center of M71 from
a 3 sec $B$  exposure taken with LRIS is shown.   
The positions of several M71 main sequence
stars in our sample in this field are indicated.\label{fig1}}
\end{figure}

\begin{figure}
\epsscale{0.7}
\plotone{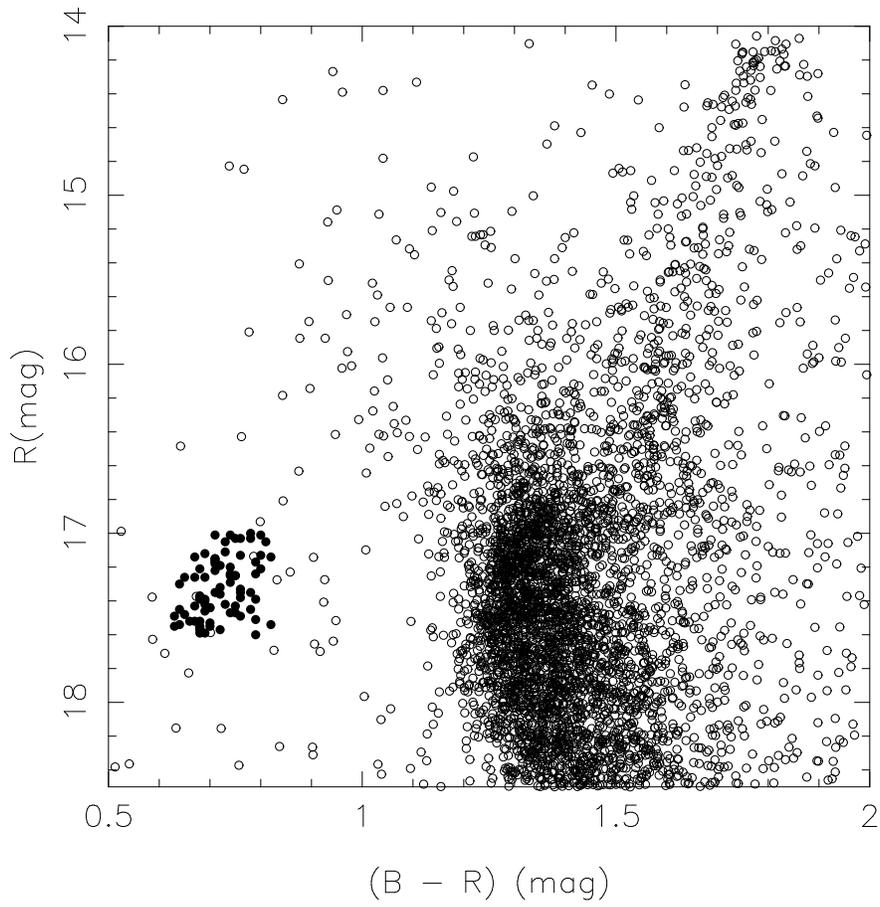}
\caption[jgcohen2.fig2.ps]{The color-magnitude diagram for the region of the
main sequence in M71.  The stars in our spectroscopic sample are shown
as filled circles and are offset 0.6 mag to the blue in color.  
\label{fig2}}
\end{figure}

\begin{figure}
\epsscale{0.7}
\plotone{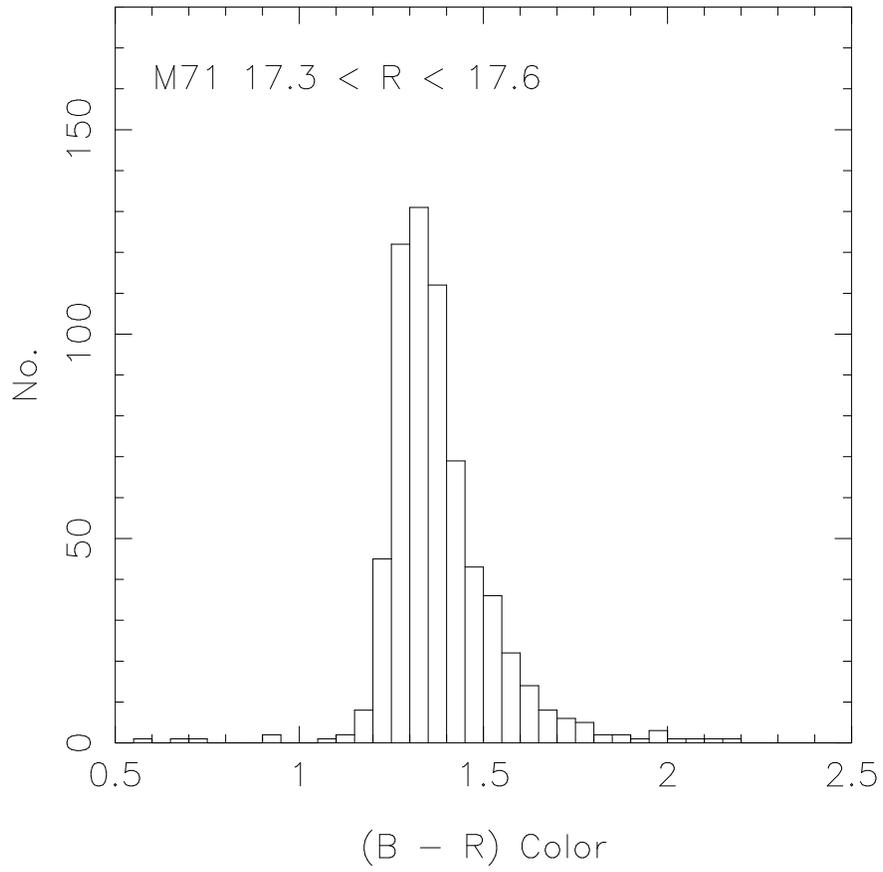}
\caption[jgcohen2.fig3.ps]{ A histogram in color of the 643 stars in the field 
of M71 with $17.3 < R < 17.6$ mag is shown.
\label{fig3}}
\end{figure}

\begin{figure}
\epsscale{0.7}
\plotone{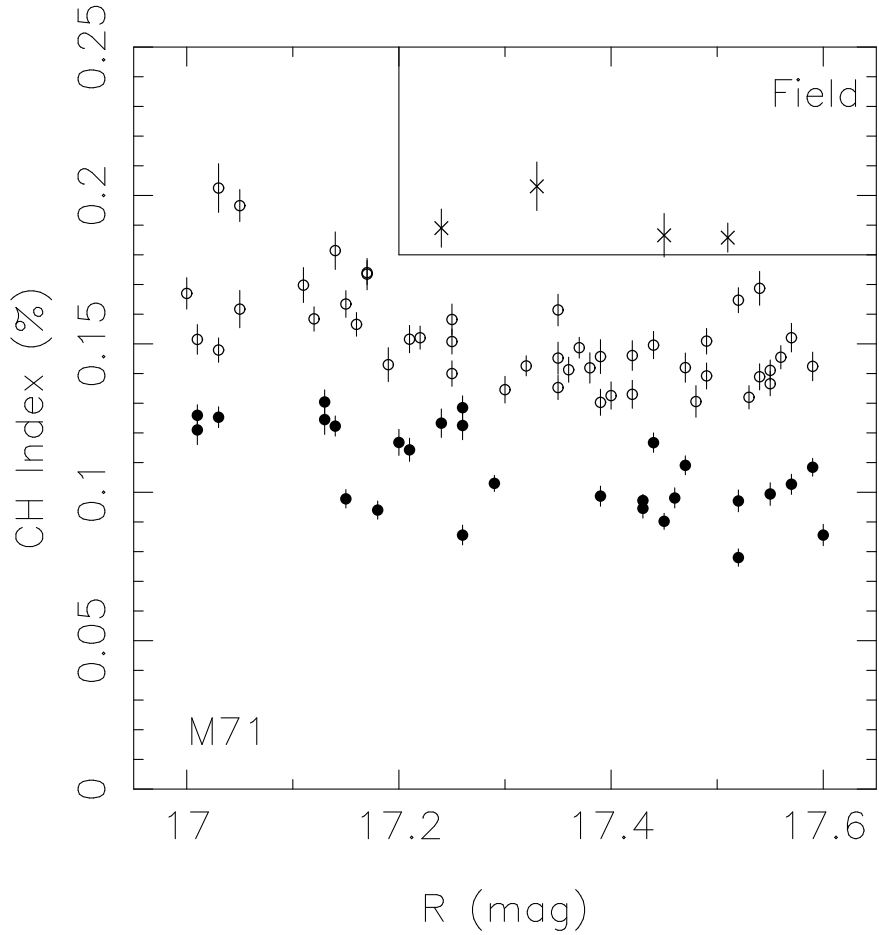}
\caption[jgcohen2.fig4.ps]{The CH indices for 79 main sequence stars in M71
are plotted as a function of $R$ mag.
The error bars shown for each point are $1\sigma$ errors calculated from
the observed count rates assuming Poisson statistics.  The symbol
coding is discussed in the text.
\label{fig4}}
\end{figure}

\begin{figure}
\epsscale{0.7}
\plotone{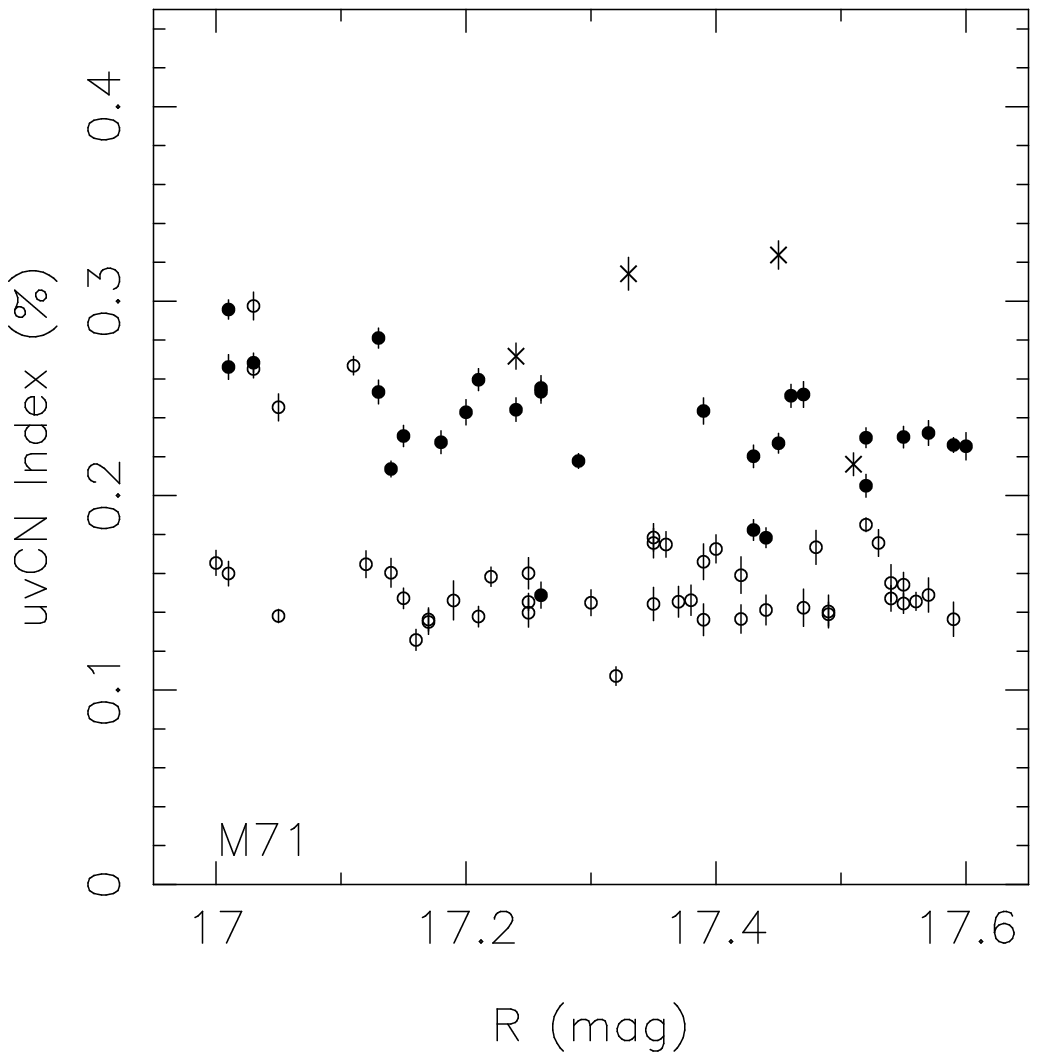}
\caption[jgcohen2.fig5.ps]{The uvCN indices for 79 main sequence stars in M71
are plotted as a function of $R$ mag.
The error bars shown for each point are $1\sigma$ errors calculated from
the observed count rates assuming Poisson statistics.  The symbol
coding is the same as that of Figure 4.
\label{fig5}}
\end{figure}


\clearpage



\begin{references}

\reference{} Briley, M.M., Hesser, J.E. \& Smith, G.H., 1994, Can. Jrl. Physics,
72, 772

\reference{} Briley, M.M. \& Cohen, J.G., 1999, manuscript in preparation

\reference{} Cannon, R.D., Croke, B.F.W., Bell, R.A., Hesser, J.E. \&
Stathakis, R.A., 1998, \mnras, 298, 601
  
\reference{} Cohen, J.~G., 1980, \apj, 241, 981
 
\reference{} Cohen, J.~G., 1983, \apj, 270, 654

\reference{} Cohen, J.~G., 1999, \aj, submitted (Paper 1)   

\reference{} Cohen, J.~G. \& Sleeper, E.~C., 1995, \aj, 109, 242

\reference{} Cohen, J.~G.,  Hogg, D.~W., Pahre, M.~A., Blandford, R.,
Shopbell, P.~L. \& Richberg, K., 1999, \apjs, in press,
Astro-ph/9809066 

\reference{} Cudworth, K.M., 1985, \aj, 90, 65 
  
\reference{} Kraft, R.P. 1994, \pasp , 106, 553
   
\reference{} Landolt, A.U., 1992, \aj, 104, 340

\reference{} Langer, G.~E., Suntzeff, N.~B. \& Kraft, R.~P., 1992, \pasp, 
104, 523

\reference{} McWilliam, A., 1997, {\it Ann.Revs.Astr.\& Astrophys.}, 35, 503   

\reference{} Oke, J.~B.,  Cohen, J.~G., Carr, M., Cromer, J., 
Dingizian, A., Harris, F.~H., Labrecque, S., Lucinio, R., Schaal, W., 
Epps, H., \& Miller, J. 1995, \pasp, 107, 307

\reference{} Pinnsoneault, M., 1997, {\it Ann.Revs.Astr.\& Astrophys.}, 35, 557

\reference{} Reed, B.C., Hesser, J.E. \& Shawl, S.J., 1988, \pasp, 100, 545

\reference{} Shortridge, K. 1988, ``The Figaro Manual Version 2.4''

\reference{} Stetson, P.B., 1987, \pasp, 99, 191
  
\reference{} Suntzeff, N.B., 1981, \apjs, 47, 1

\end{references}
\end{document}